%% file: QRNG.tex
\documentclass[%
 preprint,
superscriptaddress,
 amsmath,amssymb,
 aps,
]{revtex4-2}

\usepackage{physics}
\usepackage{graphicx}
\usepackage{subcaption}
\usepackage{dcolumn}
\usepackage{bm}
\usepackage[labelfont=bf]{caption}
\usepackage[utf8]{inputenc}
\usepackage{comment}
\usepackage{booktabs} 
\begin{document}

\preprint{APS/123-QED}

\title{40Gbps Tri-type Quantum Random Number Generator}

\author{Jiapeng Zhao}
 \email{penzhao2@cisco.com}
 \affiliation{Cisco Quantum Labs, 3232 Nebraska Ave, Santa Monica, CA, 90404, USA}
 
\author{Eneet Kaur}
 \affiliation{Cisco Quantum Labs, 3232 Nebraska Ave, Santa Monica, CA, 90404, USA}
 
\author{Michael Kilzer}
 \affiliation{Cisco Quantum Labs, 3232 Nebraska Ave, Santa Monica, CA, 90404, USA}
 
\author{Yihan Liu}
 \affiliation{Laboratory For Laser Energetics, University of Rochester, 250 E River Rd, Rochester, NY 14623, USA}

\author{Hassan Shapourian}
 \affiliation{Cisco Quantum Labs, 3232 Nebraska Ave, Santa Monica, CA, 90404, USA}
 
\author{Ramana Kompella}
 \affiliation{Cisco Quantum Labs, 3232 Nebraska Ave, Santa Monica, CA, 90404, USA}
 
\author{Reza Nejabati}
 \affiliation{Cisco Quantum Labs, 3232 Nebraska Ave, Santa Monica, CA, 90404, USA}

\date{\today}

\begin{abstract}
Traditional quantum random number generators can produce only one type of random number, while the optimal distribution of random numbers for different applications is usually distinct.
The typical solution to this challenge is either using different quantum phenomena for different types of random number, or converting one distribution of random numbers to another type. However, the former solution requires multiple hardware systems, while the latter one sacrifices a lot of secure bits. Here, we develop a quantum random number generator that can on-demand
produce three distribution types of random numbers at over 60 Gbits/s (Gbps) raw bits by measuring the quantum
vacuum noise. After randomness extraction, over 42 Gbps secure bit rate is demonstrated for uniform random numbers, and over 14 Gbps secure bit rate for Gaussian random number. Due to the lack of Rayleigh randomness extraction, only denoised Rayleigh raw bits are generated. Switching between different types of random numbers is achieved in electronics, which does not affect the generation rate. The random numbers pass NIST and Dieharder tests, and are available for various applications, which can be continuously accessed via Cisco Quantum Random Number Service.
\end{abstract}

\maketitle


\input{introduction}

\input{tri-type}
\input{experimental}

\input{randomness_extraction}

\input{Gaussian_extraction}

\input{Rayleigh_extractor}
\input{discussion}
\input{conclusion}

\section{Acknowledgement}
We appreciate the support from Gary Wang at Optoplex Corporation for his support of the balanced detection module. Jiapeng Zhao thanks for the helpful discussion with Professor Daniel Oi from the University of Strathclyde.

\bibliography{QRNG}

\end{document}

%% file: introduction.tex
\section{Introduction}
Random numbers are widely used in scientific research and engineering, including cryptography \cite{ferguson2003practical}, Monte Carlo simulations \cite{gentle2003random}, fundamental physics research \cite{shadbolt2014testing}, environmental science \cite{tucker1984numerical}, data science \cite{vershynin2018high} and artificial intelligence \cite{vovk2005algorithmic}. The most widely adopted random number generator is the pseudorandom number generator (PRNG), which generates random bits based on mathematical algorithms and secure seeds. Therefore, PRNG random numbers are predictable and deterministic once the seed is known, which can hamper the security applications of the random numbers. In addition, it is computationally hard to prove that random numbers come from a true probability distribution. \par

Quantum random number generators (QRNG) are developed to take advantage of the inherent random nature of quantum phenomena for the generation of random numbers \cite{ma2016quantum, herrero2017quantum, mannalatha2023comprehensive}. Numerous quantum processes can provide a true randomness entropy source including photon number statistics \cite{applegate2015efficient, eaton2023resolution}, spatial modes \cite{yan2014multi, meng2024generation}, temporal modes \cite{wahl2011ultrafast, nie2014practical}, vacuum noise \cite{gabriel2010generator, avesani2018source, zheng20196, drahi2020certified, huang2020gaussian, gehring2021homodyne, bruynsteen2023100}, amplified spontaneous emission \cite{nie2015generation, yang2020randomness}, laser phase noise \cite{qi2010high, xu2012ultrafast}, Raman scattering \cite{bustard2011quantum, hu2020quantum} and Bell tests \cite{liu2018device}. \par

Although various quantum entropy sources have been demonstrated for random number generation, most QRNGs can only yield one specific type of random numbers \cite{ma2016quantum, herrero2017quantum, mannalatha2023comprehensive,applegate2015efficient, eaton2023resolution, yan2014multi, meng2024generation, gabriel2010generator, avesani2018source, zheng20196, drahi2020certified, huang2020gaussian, gehring2021homodyne,  bruynsteen2023100, nie2015generation, yang2020randomness, qi2010high, xu2012ultrafast, bustard2011quantum, hu2020quantum, liu2018device, wahl2011ultrafast, nie2014practical}. However, the optimal type of random numbers for each application is usually different, and QRNGs that produce a single type of random bits may not satisfy the requirements in different application scenarios. For example, uniformly distributed random numbers are widely used in cryptography and machine learning \cite{ferguson2003practical, vovk2005algorithmic}. Gaussian random numbers are critical in financial industries and climate simulations \cite{gentle2003random}, while Rayleigh distributed random numbers are mostly used in aerodynamics and fluid mechanics \cite{tucker1984numerical, vershynin2018high}. Although multiple hardware systems can be prepared to take advantage of different quantum phenomena, this solution is not cost-effective and lacks flexibility in real applications. It is possible to convert one specific type of random number to other distributions while keeping the true randomness. However, this operation usually leads to a considerable loss of secure bits \cite{huang2020gaussian, sidhu2023finite, gryszka2021biased}. Also, during the conversion process, many approximations to well-known functions are used that can introduce numerical errors \cite{kten2011}. The proposed QRNG can on demand generate multiple types of random number distribution is desired.\par

Here, we demonstrate a tri-type QRNG that can on demand generate uniform, Gaussian and Rayleigh random bits via measuring both quadratures of the quantum vacuum noise in phase space. After removing the classical side channel information, i.e. classical correlations, via corresponding randomness extraction, we can achieve a secure random bit rate over 40 Gbits$/$s (Gbps). Since reconfiguration occurs fully in electronic postprocessing on an FPGA board, fast switching among three types of random numbers does not affect the speed of random bit generation. Our uniform random numbers also pass the NIST and Dieharder tests, and is continuously accessible via Cisco Quantum Random Number Service.\par

%% file: tri-type.tex
\section{Tri-type QRNG}
In quantum optics, measuring $Q$ or $I$ quadrature of vacuum state $\ket{0}$ via homodyne detection yields Gaussian-distributed random numbers, which has been used as the source of entropy for the generation of random numbers in previous literature \cite{gabriel2010generator, avesani2018source, zheng20196, drahi2020certified, huang2020gaussian, gehring2021homodyne, bruynsteen2023100}.\par

\begin{figure}[h]
  \centering
  \includegraphics[width=0.9\textwidth]{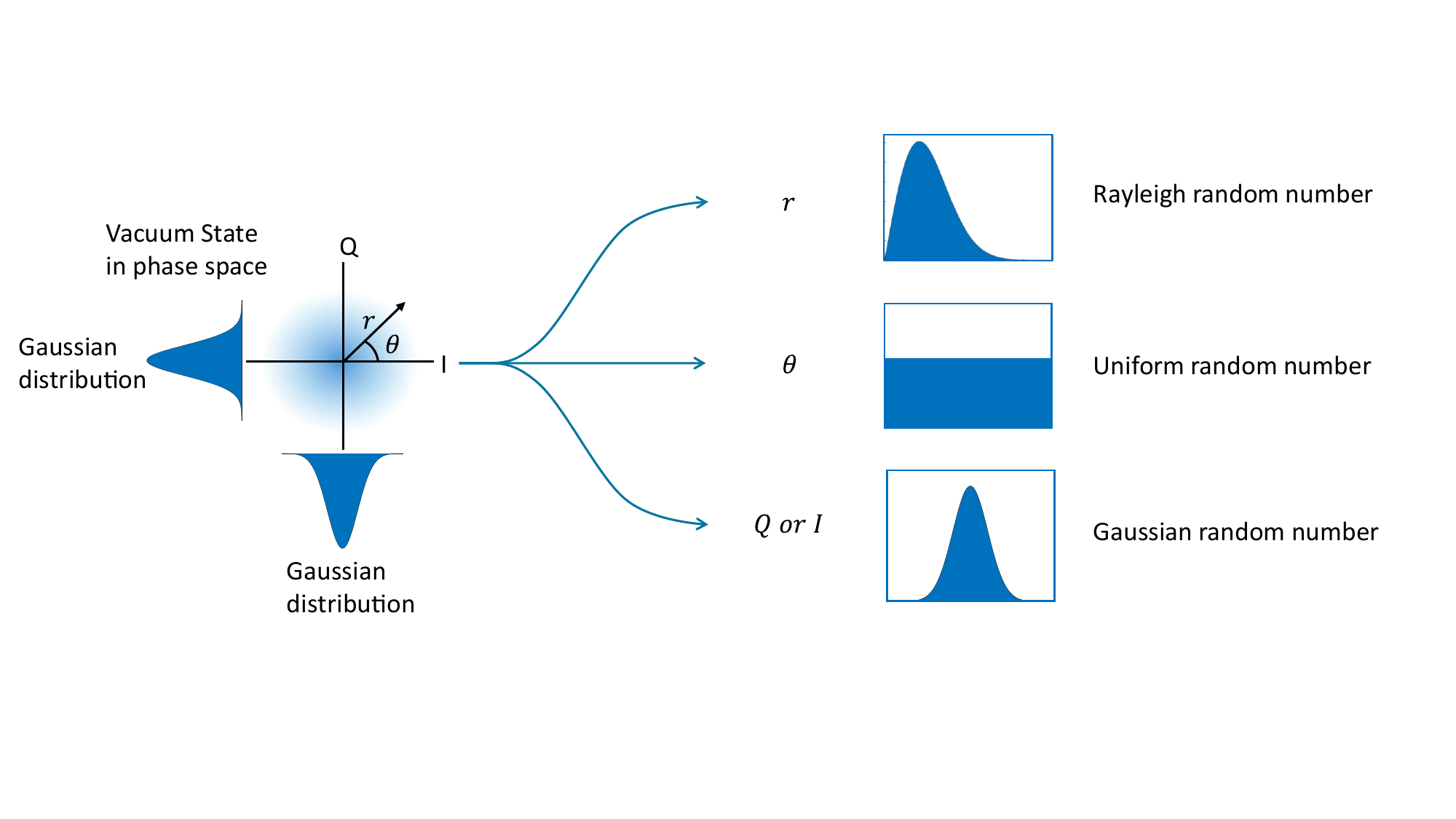}
\caption{The conceptual overview of our tri-type QRNG via dual-quadrature homodyne detection. }
\end{figure}

 Besides Gaussian-distributed random numbers, it is possible to have more types of random numbers from phase-space measurements. As shown in Fig. 1, in phase space, the phase angle $\theta$ and amplitude $r$ of the quantum vacuum state $\ket{0}$ follow the uniform distribution and the Rayleigh distribution correspondingly. Therefore,
 when we measure both the $I$ and $Q$ quadratures via dual-quadrature homodyne detection, in addition to the Gaussian distributed random number, uniform and Rayleigh distributed random numbers can be generated by calculating $\theta$ and $r$ from $I$ and $Q$. Thus, instead of measuring a single quadrature in previous single-type QRNG, our system measures both quadratures at the same time and then processes the raw bits to generate either Gaussian, uniform, or Rayleigh random bits. Since postprocessing happens solely in software, the system can accommodate multiple application scenarios, which request different types of random numbers, without modifying the hardware or converting one type of random numbers to another. As discussed later, the main limitation of the secure bit rate comes from the speed and noise performance of homodyne detectors, which is irrelevant to the type of random number generation. \par

%% file: experimental.tex
\section{Experimental realization}
The experimental configuration is shown in Fig. 2. A fiber-coupled CW 1550 nm laser (from INPHENIX) is split, by a 50:50 beam splitter (BS1), into two arms, working as the local oscillator (LO) for $I$ and $Q$ quadrature measurements. The phase of the LO laser in the lower arm is delayed by $\pi/2$ for $Q$ measurement, while the phase of the LO in the upper arm remains the same for $I$ measurement. The quantum vacuum state interferes with the LO laser at BS2 and BS3 correspondingly, and the signals are measured by two homodyne detectors (HOM1 and HOM2). After acquiring the raw bits for the $I$ and $Q$ quadratures from the transimpedance amplifiers in both homodyne detectors, the postprocessing is digitally performed on an AMD ZCU111 Radio Frequency System on Chip (RFSoC) development board, and details will be discussed later.\par

The laser has an optical power of up to 22 mW, and the linewidth is 3 kHz. BS1, BS2 and BS3 are integrated into an Optoplex 90 degree hybrid. In order to keep the LO power balanced, one variable attenuator (Thorlabs V1550PA) is inserted in each arm. It is worth noting that we do not find a significant difference in the $H_{min}$, and hence the random number rate, when we switch to a 7-Hz linewidth laser.\par

\begin{figure}[h]
    \centering
    \includegraphics[width=0.9\textwidth]{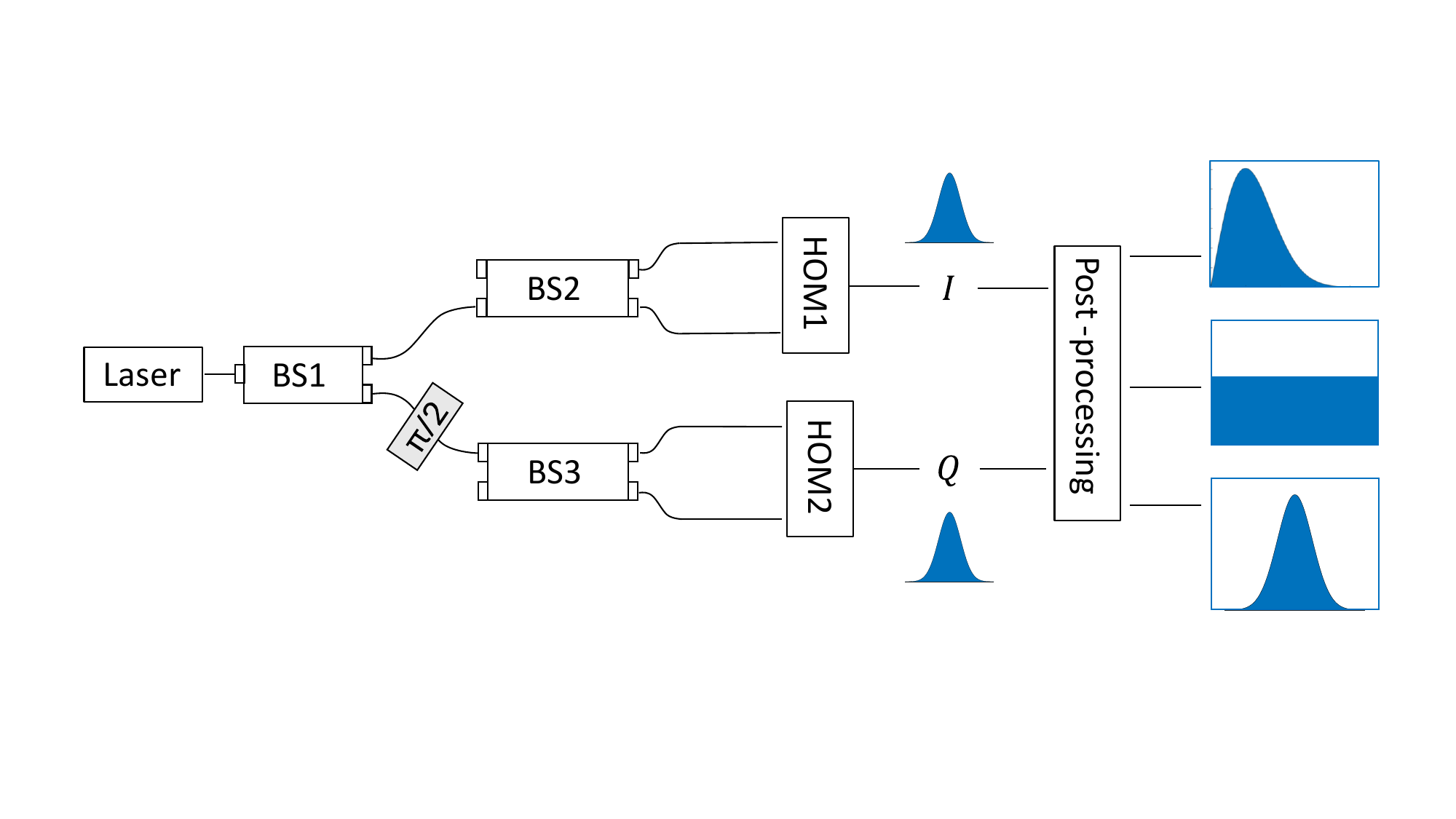}
    \caption{The schematic of the experimental realization of tri-type QRNG.}
\end{figure}

To measure quantum vacuum noise, the homodyne detector needs to be operated in the shot noise limit regime. When the homodyne detector fully rejects all classical noise, including both common-mode and non-common-mode noise, the dominant noise in measurement results would be the quantum shot noise, which is the shot noise limit regime. Because high-bandwidth homodyne detectors have a much larger noise Power Spectral Density (PSD) compared to low-speed detectors, specific optimization of the electronic circuit has to be performed to suppress the electronic noise. In order to achieve this goal, our 1.6 GHz bandwidth integrated homodyne detector module (Optoplex 90 degree hybrid with balanced receiver) is optimized by Optoplex engineers to significantly reduce the excess noise. \par

Once the dominant noise becomes quantum shot noise, the measured signal variance should have a linear dependence on the LO power. Therefore, to obtain the maximal quantum signal, the maximal optical power of LO within this linear regime needs to be determined, which can be done by subtracting the background electronic noise from the homodyne measurement noise and then performing the linear fitting of the variance under different optical power. The experimental result is shown in Fig. 3a, and the linear regression has been performed using least squares fittings with $R = 0.99$ for the LO power of 0 mW to 4.13 mW per diode. Detectors start to be saturated after 4.13 mW of LO power per diode, and the responses drop significantly when the LO power increases to 5 mW per diode. We only have one data point beyond 4.38 mW per diode because our laser has a maximal output power at 22 mW and the optical system with variable attenuators has an insertion loss of 0.99 dB. In the following discussion, all random numbers are generated at an LO power of 4.13 mW per diode, maximizing the quantum signal.\par

The PSD measurement results are shown in Fig. 3b and c. The behavior of two arms are close at frequencies below 1 GHz, with a maximal difference of 0.57 dBm at 913 MHz in homodyne signals and a maximal difference of 0.63 dBm at 702 MHz in excess noise. However, the performance difference between two detectors increases as the frequency increases, ending up with an over 4 dBm difference in excess noise at 1.6 GHz. Fortunately, the measured quantum noise in both detectors turns out to be very close until 1378 MHz, leading to a similar conditional quantum variance $\sigma_Q$.\par

\begin{figure}[h]
  \centering
  \includegraphics[width=0.9\textwidth]{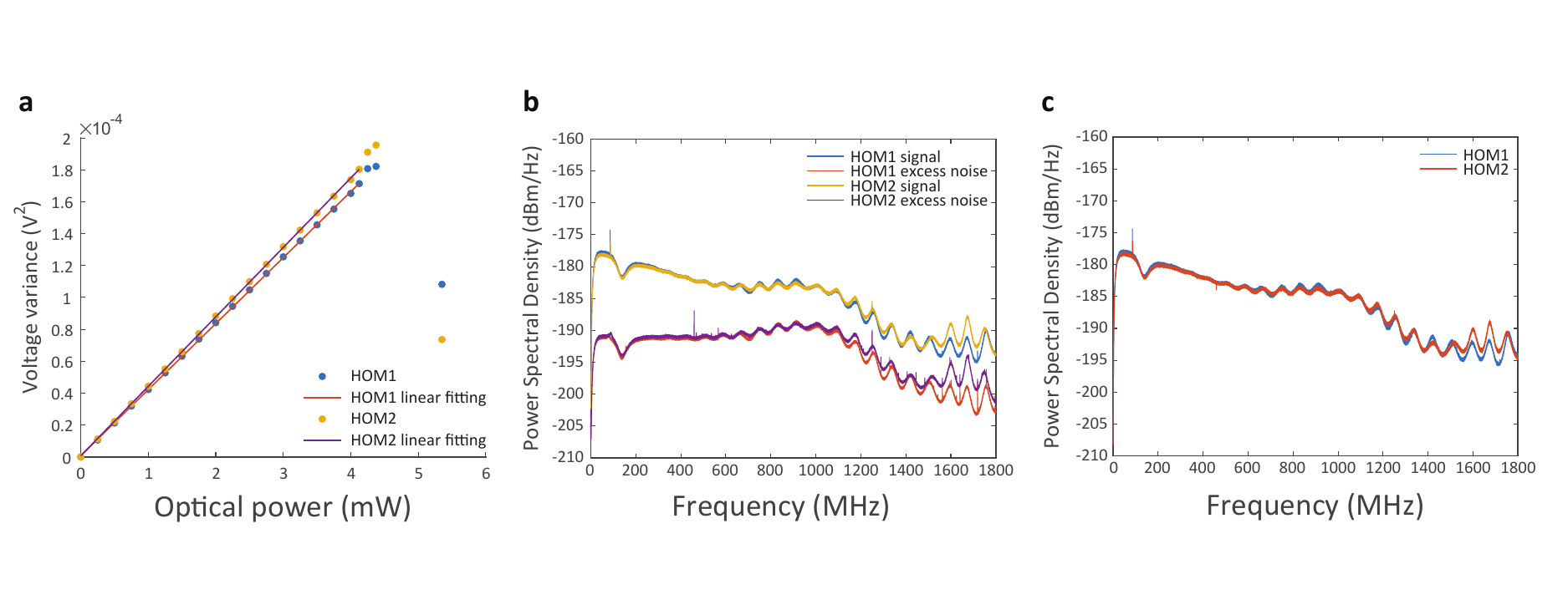}
\caption{Experimental results for quantum noise estimation. \textbf{a}. Estimate the shot noise regime of homodyne detectors. \textbf{b}. Experimental PSD measurements of homodyne signals and excess noise of homodyne detectors. \textbf{c}. Experimental measurements of quantum vacuum noise of both quadratures.}
\end{figure}

The Optoplex 90-degree hybrid provides a Gaussian-distributed analog voltage from each of its homodyne detectors via transimpedance amplifiers. The I and Q signals from the Optoplex are passed to the amplifiers to bring the analog voltage range as close to the full-scale range of the ADCs on the ZCU111 as possible. It is important to use as many ADC code bins as possible to get the widest range of random numbers from the system \cite{gehring2021homodyne, bruynsteen2023100, zhang2016fpga}. The amplified signals are then filtered with low-pass anti-aliasing filters \cite{gehring2021homodyne}.\par

The amplified and filtered voltage outputs are provided to the XM500 board (part of the ZCU111 development board kit), where the single-ended signals are converted to differential signals via baluns. The voltages are sampled by two ADCs in the RFSoC on the ZCU111 development board, and decimation filtering on the samples is also applied within the RFSoC.\par

In the programmable logic (PL) of the RFSoC, 16-bit samples are provided from the DSP datapaths of the I and Q ADC DSP \cite{AMD_PG269_2023}. The samples are queued in synchronized FIFOs to ensure sample synchronization between the I and Q channels. The user requests samples from the ADCs by toggling a register bit to begin filling the FIFOs with new sample data. The FIFOs stop filling once they are full. The user can then read out the samples containing the Gaussian-distributed sample data in the software via direct memory access (DMA). The samples can be processed via software to Gaussian and Rayleigh random numbers software.\par

The AMD PYNQ platform was used for ease of development \cite{Xilinx_RFSoC_PYNQ}. The PYNQ platform enables users to interact with the processor onboard the ZCU111 using Linux and Jupyter Notebook. The PL processing design was developed using the University of Strathclyde rfsoc qpsk design as a starting point, modifying it to collect the desired raw ADC samples and random-extracted data \cite{strath_sdr_rfsoc_qpsk}. The theory of randomness extraction and the corresponding handling of a random extractor are discussed in the next section. \par

%% file: randomness_extraction.tex
\section{Randomness Extraction}
In this section, we discuss how to extract a uniform and random signal from a noisy quantum measurement result. Let us denote the result of the measurement from a vacuum homodyne measurement using a random variable $X$. The measurement results are corrupted by various noise sources such as laser intensity and phase noise, temperature fluctuation, electronic noise, and finite range of the detectors. 

For randomness extraction, we assume that the noise is known to the eavesdropper. That is, an eavesdropper has quantum side information denoted by a random variable $E$. The length of the random sequence that can be extracted from $X$ given the quantum side information $E$ is bounded by the leftover hash lemma \cite{Tomamichel_2011}:
\begin{equation}
    l \geq n H_{\textrm{min}}(X|E) - \log \frac{1}{2\epsilon^2}, \label{eq:hash_bound}
\end{equation}
where $n$ is the length of the input bit string, $\epsilon$ is the security parameter, and $H_{\textrm{min}}(X|E)$ is the min-entropy.

Next, we calculate $H_{\textrm{min}}(X|E)$ for the experimental setup described above and provide a procedure to extract a random signal from $X$. 

Previous work \cite{gehring2021homodyne,bruynsteen2023100} has characterized the quantum side information available to an eavesdropper, providing a lower bound on the min-entropy. In the following, we summarize the method for clarity.

First, assume an independent and identical Gaussian noise model. The side information is assumed to arise from excess noise. The min-entropy can then be bounded as \cite{gehring2021homodyne}:
\begin{equation}
    H_{\textrm{min}}(\overline{X}|E) \geq -\log \left[(\sqrt{n} + \sqrt{1+n})^2 \cdot \erf\left(\frac{\Delta x}{2 g_\star}\right)\right],
\end{equation}
where $g_\star$ is chosen such that:
\begin{equation}
    \erf\left(\frac{\Delta x}{2 g_\star}\right) = \operatorname{erfc}\left(\frac{R}{g_\star}\right).
\end{equation}

Here, $n$ is the mean number of photons in the thermal state, $R$ is the ADC range, and $\Delta x$ is the size of the ADC bin.

For nonlinear ADCs, where the bin size $\Delta x$ varies due to nonlinearity, the min-entropy bound becomes:
\begin{equation}
   H_{\textrm{min}}(\overline{X}|E) \geq -\log\left[\Gamma(n) \cdot \erf\left(\frac{R/2^N + \text{DNL}_{\text{max}}}{2 g_\star}\right)\right], \label{eq:min-entropy}
\end{equation}
where $\text{DNL}_{\text{max}}$ captures the nonlinearity of the ADC.

To account for the non-iid nature of noise due to the finite bandwidth of the detector, we use the approach in \cite{bruynsteen2023100}, which introduces an effective iid model. The effective mean photon number $n$ is given by:
\begin{equation*}
    n = \frac{1}{2}\frac{\sigma_M^2}{\sigma_{Q,c}^2} - \frac{1}{2},
\end{equation*}
where $\sigma_M^2$ is the variance of the measured signal, and $\sigma_{Q,c}^2$ is the conditional variance of the quantum signal $Q$, calculated from the power spectral density. Substituting the effective $n$ into Eq.~\ref{eq:min-entropy}, we obtain a lower bound on the min-entropy that handles the non-iid noise. See \cite{bruynsteen2023100} for details.

In this section, we have described a method to lower bound the min-entropy. Substituting this bound into Eq.~\ref{eq:hash_bound}, we can determine the length of the random sequence extractable from the measurement result. In next section, we will describe how to implement randomness extractor for raw quantum bits following different distributions.

\section{Randomness Extractor}
A randomness extractor maps a weakly random string to a nearly perfect $\varepsilon$ random string, where the distribution is uniform and random ($0 < \varepsilon < 1$). We use two types of extractors: the Toeplitz extractor and the Dodis extractor \cite{dodis2004improved}. The Dodis extractor provides a seed for the Toeplitz extractor. The implementation of the Dodis extractor is based on Appendix D of \cite{ferguson2003practical}.

\subsection{Toeplitz Extractor}
The Toeplitz extractor is commonly used for amplification of privacy. In \cite{ma2013postprocessing}, the authors proved that the Toeplitz extractor, based on the universal hashing function \cite{wegman1979new}, is a strong extractor by means of the Leftover Hash Lemma \cite{Tomamichel_2011}. This implies that the seed used in the Toeplitz matrix can be reused in subsequent runs. Therefore, even though the seed for the Toeplitz matrix is of length $n + m - 1$, the ability to reuse the seed ensures net randomness extraction, where $n$ is the input size and $m$ is the output size.

To implement the Toeplitz extractor, follow these steps:

\begin{itemize}
    \item Given an initial sequence of size $n$, min-entropy $H_{\text{min}}$, and security parameter $\varepsilon$, set $m = nH_{\text{min}} + 2 \log \varepsilon$.
    \item Construct an $n \times m$ Toeplitz matrix using $n + m - 1$ random seed bits. Specify the first row and column using the random seed, with other elements determined by moving the rows and columns diagonally.
    \item Multiply the weak $n$ bit random sequence by the Toeplitz matrix to obtain the final random sequence.
\end{itemize}


To generate the $n - m + 1$-bit seed required, we use the output of the two-seeded Dodis extractor, as implemented in \cite{foreman2023practical}. 

In the experiment, after the samples are processed by RFSoC, the programmable logic routes the samples from each FIFO to a Toeplitz randomness extraction block. The block in each channel is based on designs described in the previous literature \cite{zheng20196,zhang2016fpga, guo2024parallel}. The blocks take in 1536 input bits of noisy Gaussian samples from the ADCs and output 1024 bits per extraction. The uniform random bits from the randomness extractor blocks are provided to DMA for access by the software on-board the RFSoC. Data can also be used on board, stored or sent elsewhere for later use.\par

For uniform random number extraction, the system operates with an effective sample rate of 2 GHz, leading to 42.66 Gbps of extracted uniform random bits in total, instantaneously. The sample rate is slower than the 3.2 GHz sample rate, limited by detector bandwidth, due to the lack of a high-speed, low-noise anti-aliasing filter. Future work could increase the sample rate to 3.2 GHz and hence extract a random bit rate over 68.25 Gbps. \par

%% file: Gaussian_extraction.tex
\subsection{Gaussian extractor}

To remove excess classical noise and address imperfections in the digitization process, effective randomness extraction for Gaussian raw bits is crucial. In this work, we adopt a modified recursive method based on the Wallace method to directly extract Gaussian distributed data from raw input~\cite{huang2020gaussian}. This approach aims to minimize classical noise and digitization artifacts influence, ensuring a high-quality Gaussian output suitable for simulation and modeling application.

\paragraph{Stage 1: Most Significant Bit (MSB) Selection}
The first step employs an entropy-based truncation strategy. For each 16-bit ADC sample, we retain only the $m$ most significant bits (MSBs), where $m$ is determined based on the estimation of the conidtional min-entropy $H_{min}$. This selection is grounded based on that classical noise typically fluctuates within a narrow range, predominantly affects the least significant bits (LSBs), whereas quantum shot noise more strongly influences the MSB region. By discarding LSBs, we effectively suppress classical noise components without significant loss of quantum entropy. \par

\paragraph{Stage 2: Modified Recursive Matrix Method}
To further enhance the Gaussian distribution and de-correlate residual structure in MSB-truncated samples, we applied the modified recursive matrix method. The basic principle is that the summation of Gaussian-distributed variables remains Gaussian. By grouping data and applying a modified recursive matrix, this method directly produces a new set of Gaussian-distributed data. It builds on the original recursive method, commonly known as the Wallace method. 

\begin{figure}[h!]
\centering
\includegraphics[width=0.8\textwidth]{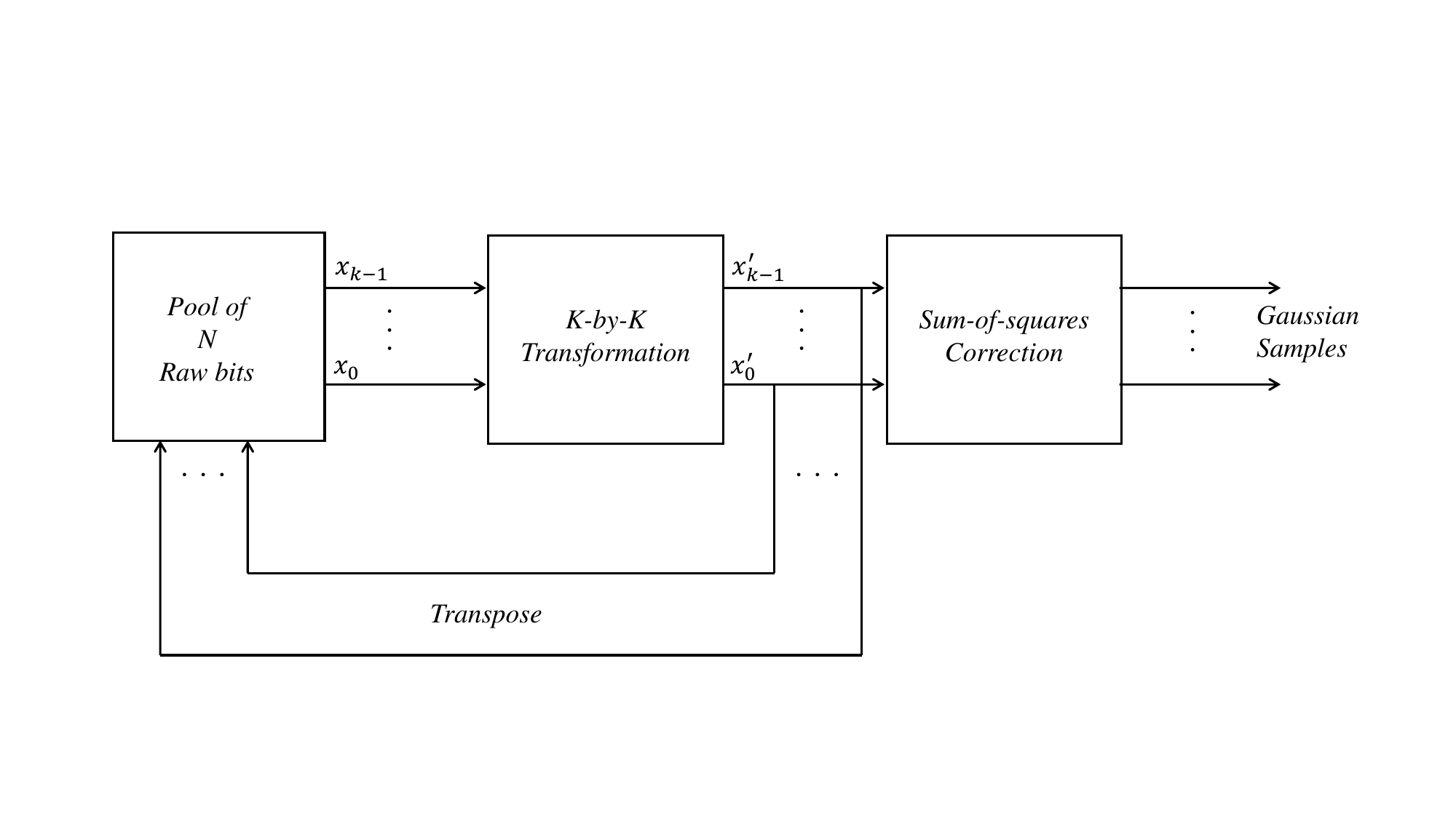}
\caption{Modified recursive method work flow}
\label{fig:workflow}
\end{figure}

An overview of the modified recursive method is illustrated in Fig.~\ref{fig:workflow}. It takes a pool of Gaussian-distributed raw numbers, with a total of \(N = K \cdot L\) samples. Here, each transformation step operates on groups of \(k\) samples, and a total of \(L\) such groups are processed per pass. These samples are first normalized to ensure that the average squared value is equal to one. The input data points \(N\) are then divided into smaller groups of size \(k\), with each group treated as a vector. Subsequently, each vector is multiplied by a modified recursive matrix of dimensions \(k \times k\). This matrix is specifically designed to transform the data, producing a new set of data with extended precision~\cite{huang2020gaussian}. Given that there are \(K \cdot L\) variables in the raw data pool, \(L\) transformation steps are performed during each pass.

Following the original Wallace method, the transformed data undergo additional mixing. During this process, the data vector transitions from row-major to column-major order, involving a transpose operation in each iteration. This reorganization ensures that, after multiple passes, every value in the pool contributes to every value in the transformed pool. To maintain consistency, a sum-of-squares correction is applied at the end of each step, preserving the overall summation of the data throughout the process.

During entropy estimation in the post-processing stage of the experiment, the ADC utilized in the experiment operates with a sampling range of \(128 \, \text{mV}\) peak-to-peak and a sampling precision of 16 bits. This configuration yields a min-entropy of \(H_{\text{min}} = 0.70\) for the $I$ quadrature and \(H_{\text{min}} = 0.71\) for $Q$ quadrature. 

In the recursive method, with \(K = 4\), each step processes four data points. The output precision is calculated as \(n = m + K - 1 = 14\) bits per signal, where \(m\) denotes the min-entropy of the input data. This process ensures that the transformation preserves key statistical properties while extending the precision of the resulting signals.

\begin{figure}[h!]
\centering
\includegraphics[width=0.8\textwidth]{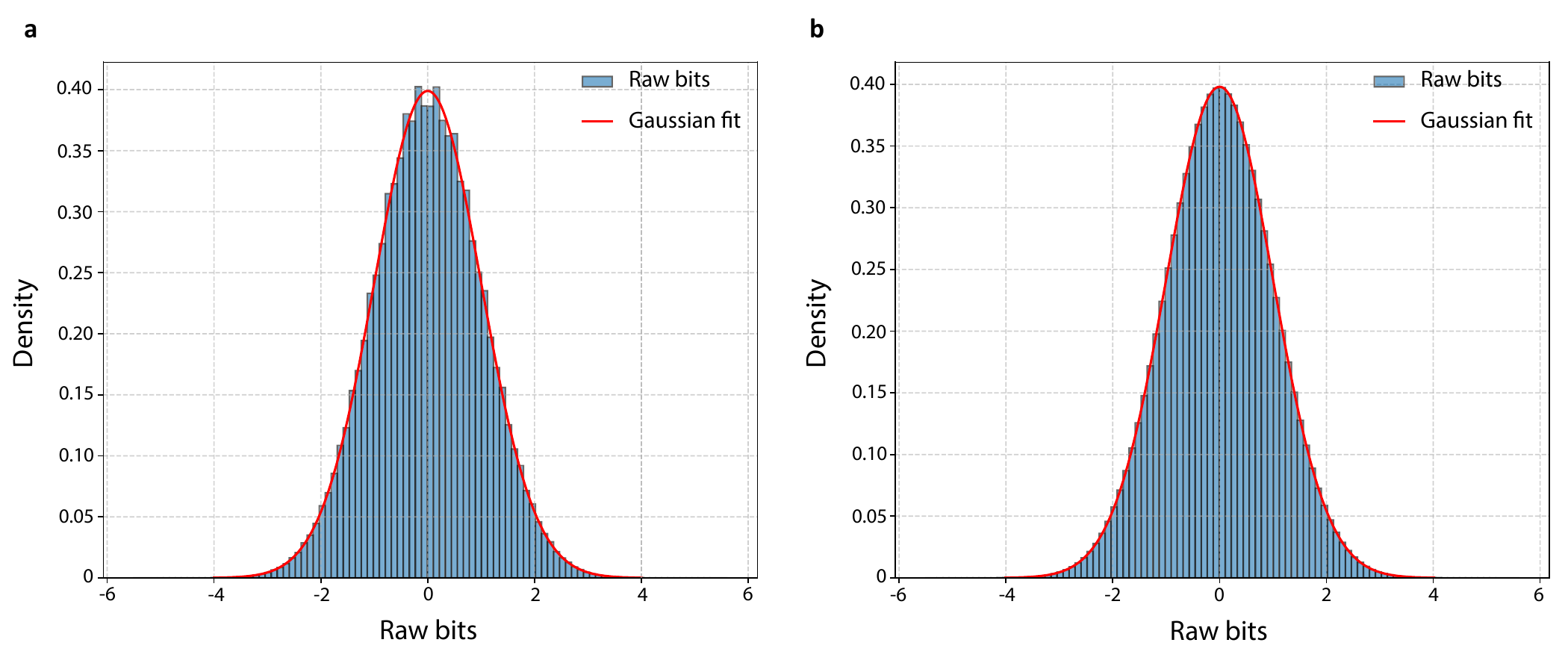}
\caption{Experimental histogram of Gaussian-distributed bits for $Q$ quadrature before and after randomness extraction.}
\label{fig:histograms_gaussian_ch2}
\end{figure}

The test results for one of the two channels in the experiment are shown in Fig.~\ref{fig:histograms_gaussian_ch2}. Detailed goodness-of-fit (GoF) test data for both channels are presented in Table~\ref{tab:gof_test}. The results indicate that, after five rounds of operations in $I$ and four rounds in $Q$, both channels passed the GoF tests, producing real Gaussian distributed data. \par 

In theory, it is possible to extract up to 0.7 bits of Gaussian random bit from 1 bit of raw Gaussian distributed bit. However, due to the low-efficient extractor, the extracted Gaussian random number rate is measured to be 14.01 Gbps. A more efficient Gaussian random number extractor needs to be designed.

\begin{table}[h!]
\centering
\caption{Goodness-of-Fit (GoF) Test Results for $I$ and $Q$}
\label{tab:gof_test}
\begin{tabular}{lcccc}
\toprule
\textbf{Test}       & \textbf{$I$  Raw Bits} & \textbf{$I$ Gaussian Bits} & \textbf{$Q$ Raw Bits} & \textbf{$Q$ Gaussian Bits} \\ 
\midrule
KS-test             & $p = 9.756 \times 10^{-121}$ & $p = 0.7426$       & $p = 2.622 \times 10^{-138}$ & $p = 0.2661$        \\ 
Chi-Squared test    & $p = 0$                     & $p = 0.4971$       & $p = 0$                     & $p = 0.7295$        \\ 
\bottomrule
\end{tabular}
\end{table}

Although the empirical results confirm that the extracted data pass the Gaussian GoF tests, this method lacks theoretical proof of randomness preservation and is thus not entirely reliable. The recursive nature and data reuse may introduce subtle correlations, although negligible for simulation use cases, that limit the extractor's application in cryptographic contexts. This aligns with previous concerns raised in the original Wallace method~\cite{wallace1996fast}. \par

%% file: Rayleigh_extractor.tex
\subsection{Rayleigh extractor}
Using the same dataset, Rayleigh-distributed raw bits can be generated by calculating the square root of the sum of the squared values from $I$ and $Q$ data. The raw bits fail the Chi-squared goodness-of-fit (GoF) test, yielding a \(p = 0\). Unfortunately there is no quantum random number extractor for Raleigh distribution yet, and several approaches have been explored to address this challenge here.

\begin{figure}[h!]
    \centering
    \includegraphics[width=0.8\textwidth]{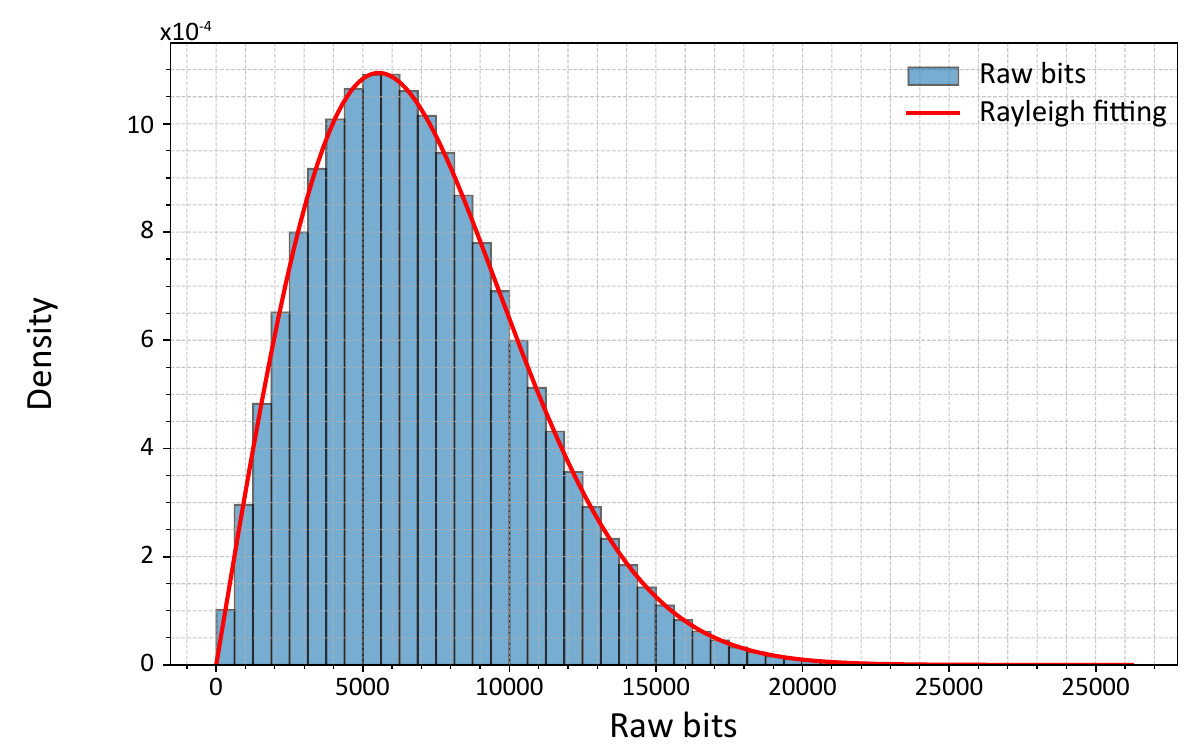} 
    \caption{Experimental histogram data, processed by Savitzky-Golay filter, to the Rayleigh distribution.}
    \label{fig:rayleigh_distribution}
\end{figure}

Based on insights from the Gaussian extractor design, the development of a Rayleigh quantum number extractor requires significant advances in both hardware and software optimization. Hardware improvements have greatly enhanced extraction efficiency, reducing the number of rounds required by the modified recursive method in the Gaussian extractor. Simultaneously, test results indicate that noise levels in the raw Gaussian and Rayleigh distributions have been substantially reduced. Besides, similar to the development of a Gaussian randomness extractor, a dedicated Rayleigh randomness extractor can be developed on the basis of classical filters. We have tested and modified a few classical filters for Rayleigh raw bits, including Wiener filter, Savitzky-Golay filter and wavelet denoise filter. Although the observed noise level can be reduced after these filters, only the Savitzky-Golay filter shows a slight improvement in the GoF test results, which is shown in Fig.~\ref{fig:rayleigh_distribution}.  These trials demonstrate the potential of combining hardware and software optimization techniques to address classical noise in quantum signals, and the foundation of developing a rigorous Rayleigh randomness extractor in the future. However, at present, a method to digitally remove all classical correlations and extract Rayleigh distributed quantum random number from raw Rayleigh bits has not been identified.


%% file: discussion.tex
\section{Discussion}
Our random numbers, generated by the hardware system in real time, can be accessed as a cloud service through Cisco Quantum Random Generator page\cite{cisco2025qrng}. The system described above continuously (24 hours per day, 7 days per week) generates secure bits on the FPGA board by monitoring the system noise. After real-time testing of the secure numbers, numbers that fully pass all tests are uploaded to Cisco Cloud for users, which eliminates the possible environmental-dependent noise in random bits generation. 

Most applications requiring Gaussian random numbers, such as Monte Carlo simulations and probabilistic modeling, do not need cryptographic-level security of random numbers. For such applications, Gaussian distributed random numbers that are statistically indistinguishable from the theoretical distribution are enough for better modeling. Therefore, our method offers a practical and high-speed solution for real-time Gaussian QRNG with a demonstrated secure bit rate of 14.01 Gbps.

With regard to Rayleigh quantum random numbers, a randomness extractor needs to be developed. Our experimental observation shows that classical filters can improve the quality of random numbers but cannot pass statistical tests, indicating that a future Rayleigh extractor might be developed based on classical filters like Gaussian randomness extractor.

%% file: conclusion.tex
\section{Conclusion}
We propose and experimentally demonstrate the first QRNG system that generates on demand three different types of random numbers with the same hardware. The extracted uniform random number rate is over 42 Gbps, with the potential to reach 68 Gbps with a better low-noise filter. After implementing the Gaussian random number extractor, more than 14 Gbps Gaussian random numbers are extracted, and the rate can be improved to 22 Gbps after the implementation of a high-bandwidth low-noise filter. The Gaussian random number generation rate is mainly limited by the inefficient randomness extractor, and a higher theoretical rate can be achieved by a carefully designed Gaussian extractor. Although the current implementation does not yet provide a rigorous extractor for Rayleigh random numbers, these improvements in the raw Rayleigh bits show a strong foundation for future development. Our system provides a tangible solution to provide different types of quantum random numbers to accommodate various applications.